# Directional waveguide coupling from a wavelength-scale deformed microdisk laser

Brandon Redding[1], Li Ge[2], Glenn S. Solomon[3], and Hui Cao[1]

[1]Department of Applied Physics, Yale University, New Haven, CT 06511, USA
[2]Department of Electrical Engineering, Princeton University, Princeton, NJ 08544, USA
[3]Joint Quantum Institute, NIST and University of Maryland, Gaithersburg, Maryland 20899, USA

We demonstrate uni-directional evanescent coupling of lasing emission from a wavelength-scale deformed microdisk to a waveguide. This is attributed to the Goos-Hänchen shift and Fresnel filtering effect that result in a spatial separation of the clockwise (CW) and counter-clockwise (CCW) propagating ray orbits. By placing the waveguide tangentially at different locations to the cavity boundary, we may selectively couple the CW (CCW) wave out, leaving the CCW (CW) wave inside the cavity, which also reduces the spatial hole burning effect. The device geometry is optimized with a full-wave simulation tool, and the lasing behavior and directional coupling are confirmed experimentally.

Over the last two decades there have been tremendous developments in the field of microlasers. Recent advances in nanophotonics and integrated optics have led to a strong interest in further reduction of the laser size. Nanoscale coherent light sources have been proposed and were recently demonstrated[1-4]. Fabrication of such devices is technically challenging and the device performance usually deteriorates with size reduction, e.g. the lasing threshold increases and the quantum efficiency decreases. Besides a smaller footprint and tighter confinement of light, are there any advantages to using wavelength-scale or even smaller lasers? In this paper, we address this question using a waveguide-coupled disk laser as an example. We will show that the size reduction provides an opportunity to achieve novel functionalities which cannot be realized in larger devices.

Semiconductor microdisk lasers are excellent candidates for on-chip light sources in integrated photonics applications due to their simple design geometry, small device size, low lasing threshold, and ability to couple output power to planar integrated waveguides[5,6]. Strong optical confinement is provided by total internal reflection of light from the disk boundary, resulting in very high quality ($Q$) factors and low lasing thresholds. Laser emission can be evanescently coupled to a waveguide placed close to the disk boundary along the tangential direction. The chiral symmetry of a microdisk ensures that all lasing modes contain equal amplitudes of clockwise (CW) and counter-clockwise (CCW) propagating waves, thus the energy coupled out is split evenly into two waveguide directions. One way to obtain uni-directional coupling is to break the chiral symmetry, e.g., in a spiral shaped cavity[7,8]; another approach is to place a waveguide along the normal direction of the cavity boundary[9]. In all of these studies, the cavity size was much larger than the emission wavelength.

In this work, we demonstrate that a *wavelength-scale* deformed disk laser can produce directional output to a waveguide placed tangentially to its boundary. The chiral symmetry is broken not by the deformation of a circular cavity shape, but by the Goos-Hänchen shift (GHS) and the Fresnel filtering (FF) effect that become significant when the cavity size is comparable to the wavelength. GHS describes a lateral displacement (on the order of a wavelength) of a totally internally reflected beam, and FF has the effect of deflecting the reflected beam away from the specular reflection direction[10-12]. In wavelength-scale cavities without rotational symmetry, these effects serve to split a periodic ray orbit into two pseudo orbits (with equal length), one propagating CW and the other CCW[13]. The spatial separation of CW and CCW orbits allows us to couple one of them more efficiently than the other by placing the waveguide at a specific location on the cavity boundary. By moving the waveguide to another location on the boundary, we can couple the other orbit out. Selective coupling of one orbit reduces the amplitude of this orbit inside the cavity, making the other orbit dominant. In other words, with this selective coupling scheme, the lasing mode is composed mainly of the less-coupled CW or CCW orbit. The standing wave pattern is thus replaced by a propagating wave, which produces a more uniform spatial distribution of the field intensity inside the cavity. Hence, the spatial hole burning effect is reduced, and the lasing mode can utilize the optical gain that is not accessible to a standing wave pattern because it is at the field nodes.

Recently, we demonstrated the breaking of chiral symmetry between CW and CCW propagating waves by GHS and FF for low-Q modes in a deformed cavity [13]. In this paper, we consider the same cavity shape but increase the wavelength (equivalent to reducing the cavity size) so that the GHS and FF effect become dominant for the high-Q modes. . The cavity boundary can be described in polar coordinates as $\rho(\phi) = R [1 + \varepsilon \cos(\phi)] [1 - \varepsilon_1 \cos(2\phi)] + d$, where $R$ = 890 nm, $\varepsilon$ = 0.28, $\varepsilon_1$ = 0.06 and $d$ = 60 nm. We numerically simulate the resonant modes of transverse electric (TE) polarization (electric field in the plane of the disk) using the finite element method. The refractive index $n$ of the disk is set to 3.13. We first consider the high-Q mode with azimuthal number $m$=10 ($\lambda$ = 1355 nm, $kR$ = 4.13). Our simulations reveal that this is a scar mode localized on an unstable periodic orbit with three bounces along the cavity boundary. The Husimi analysis of this mode (similar to that in Ref. [13]) illustrates that the triangle orbit is split into CW and CCW pseudo orbits due to the GHS and FF effect, as shown in Fig. 1(a). The spatial separation of the CW and CCW orbits provides an opportunity to selectively couple light from one of them by positioning a waveguide tangential to a part of the cavity boundary that is closer to one of the orbits. To optimize this selective coupling, we introduced a straight waveguide separated from the cavity boundary by 100 nm, and varied the location of the coupling point along the boundary. In our numerical simulations, the waveguide (having the same refractive index as the disk) was oriented vertically to the right of the disk, and the disk was rotated in the CW direction (Fig. 1). The angle of rotation, $\theta$, specifies the coupling point. At each $\theta$, we simulated the intensity of emission coupled to the waveguide in the CW and CCW directions. In Fig. 1(d), we compile these results by plotting the percentage of the intensity coupled in either the CW or CCW directions as a function of $\theta$. At $\theta$ = 90°, 70% of the emission is coupled in the CCW direction, while at $\theta$ = 150°, 80% of the emission is coupled in the CCW direction [Fig. 1(b,c)]. The results presented above for a particular mode are general and hold for other modes with similar values of $kR$. However, for modes with much larger $kR$, e.g. the one with $m$ = 25, we cannot obtain directional coupling. This is because the GHS and FF effect are significantly reduced, at least for the high-Q modes. In other words, the directional coupling effect diminishes for large cavities.

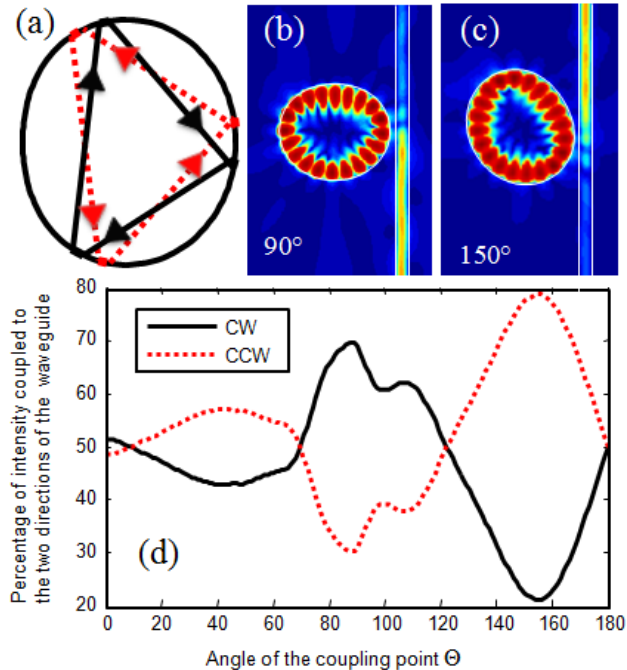

**Fig. 1.** Numerical simulations of the waveguide coupled deformed microdisk. (a) The CW (black solid line) and CCW (red dotted line) pseudo-orbits of the $m$=10 mode are spatially separated by the GHS and FF effects. (b, c) Spatial distribution of the $H_z$ field amplitude showing directional coupling to the waveguide with the cavity rotated 90° or 150° relative to the waveguide. (d) Percentage of the intensity coupled in the CW (black solid line) and CCW (red dotted line) directions as a function of the angle $\theta$ that the cavity is rotated relative to the waveguide.

In the presence of the waveguide, the resonant mode in the cavity no longer contains equal amounts of CW and CCW waves, and its standing wave pattern disappears. From the mode profiles in Fig. 1 (b,c), we no longer see the nulls in the field amplitude that exist in the standing wave pattern without a waveguide. For a quantitative comparison, we plot in Fig. 2(a) the intracavity field amplitude along the azimuthal direction with a waveguide at $\theta$ = 150° and without a waveguide. The modulation is dramatically reduced in the presence of the waveguide and the nulls disappear. We also performed a Bessel decomposition to extract the azimuthal components of the CW and CCW waves for this mode, with and without a waveguide. As shown in Fig. 2(b), in the absence of a waveguide, the CW ($m < 0$) and CCW ($m > 0$) waves have equal amplitudes; however, when the waveguide is introduced, the CW wave becomes dominant. Note that in this coupling configuration, it is primarily the CCW wave that couples to the waveguide. Figure 2(b) reveals that the dominant component remaining inside the cavity is the CW wave. The removal of the standing wave pattern for a lasing mode reduces the spatial hole burning effect at the locations of field maxima and allows the utilization of the optical gain at the locations of field nulls.

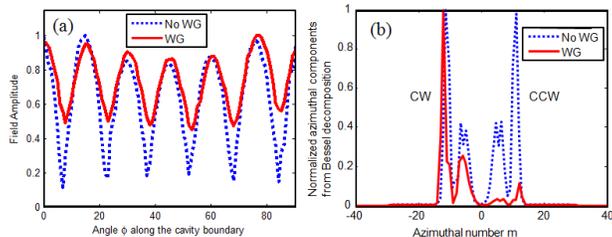

**Fig. 2.** (a) Calculated intracavity field amplitude along the azimuthal direction with a waveguide (WG) at $\theta$ = 150° (red solid line) and without a waveguide (blue dashed line). The field modulation is reduced in the presence of a waveguide. (b) Bessel decomposition of the azimuthal components in the CW and CCW directions for the same modes. In the presence of a waveguide, the CCW components couple to the waveguide and the remaining CW components dominate the mode in the cavity.

To verify these behaviors experimentally, we fabricated the waveguide-coupled wavelength-scale cavities with GaAs using InAs quantum dots as the gain material. The sample was grown epitaxially on a GaAs substrate via molecular-beam epitaxy. The structure consisted of a 1 μm $Al_{0.68}Ga_{0.32}As$ layer and a 260 nm GaAs layer with five embedded layers of InAs quantum dots. The cavity and coupling waveguide were patterned via electron-beam lithography. The pattern was transferred to the semiconductor via inductively coupled plasma etching with a $BCl_3$ and $Cl_2$ mixture. The underlying $Al_{0.68}Ga_{0.32}As$ layer was then selectively etched with hydrofluoric acid to provide better optical confinement in the GaAs layer. The disk shape, shown in the high resolution scanning electron micrographs (SEM) in Fig. 3(a,b), matches well with the designed one. Note that the device size was scaled down to R = 700 nm so that the mode wavelength ($\lambda$) falls in the emission regime of the InAs quantum dots (~900 nm). The waveguides were 6 μm long, terminated by large tapered regions shown in Fig. 3(c). The tapered regions provide a support for the waveguide which is free-standing in the coupling region near the disk after the undercut. The rough background in the SEM is the residual AlGaAs material left on the GaAs substrate after the selective wet etch.

The devices were tested in a liquid helium cryostat at 10° K. Optical excitation was provided by a mode-locked Ti:Sapphire laser ($\lambda$ = 790 nm) operating at 76 MHz with 200 fs pulses. The pump beam was focused onto the surface of a single disk with a 50× long working distance objective lens. We observed lasing in three disks, each having the same size and shape, but different orientation relative to the coupling waveguide. The lasing modes in all three disks correspond to the high-$Q$ mode with $m$ = 12 and $\lambda$ ~905 nm [inset of Fig. 4(a)]. The small difference (less than 1%) in $\lambda$ results from tiny variations in the disk size/shape that were introduced during the fabrication process. As shown in Fig. 4(a), the dependence of the emission intensity on the pump power exhibits a threshold behavior, confirming the onset of lasing action.

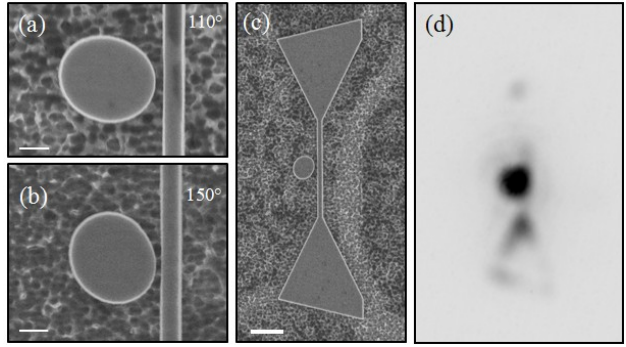

**Fig. 3.** SEM images of the deformed microdisk cavities rotated 110° (a) and 150° (b) relative to the coupling waveguides. The scale bars are 500 nm. (c) SEM image of the entire device. The tapered triangular regions support the suspended waveguide. The bottom and top edge of the tapered regions were angled to avoid light reflecting back into the waveguide. The scale bar is 2 µm. (d) Image of the laser emission showing directional coupling.

The laser emission that is coupled from the disk to the waveguide propagates along the waveguide until it reaches one of the tapered regions. Some of the emission is then scattered out of the plane, and collected by the same objective lens that focused the pump light. The collected emission is imaged onto a charge coupled device (CCD) camera by another lens. A narrow-band interference filter was positioned in front of the camera to select the laser emission near 905 nm. Figure 3(d) shows an optical image taken by the CCD camera of the deformed microdisk rotated 150° relative to the waveguide. A strong signal is registered from the microdisk itself as well as scattered light from the triangular pedestal at the bottom of the image. A much weaker scattering is visible from the pedestal at the top of the image, indicating that the majority of the lasing emission coupled out in the CW direction. In order to quantify the percentage of the emission which coupled in the CW or CCW direction, we integrated the scattered intensity over the bottom or top pedestal. In Fig. 4(b), we overlay the experimental data points ("×") for the three lasing devices on the curves of simulated coupling percentages for the $m = 12$ mode. The experimentally measured coupling efficiencies agree well with the predictions, and a maximum coupling efficiency of nearly 80% is observed from the device with $\theta = 150°$.

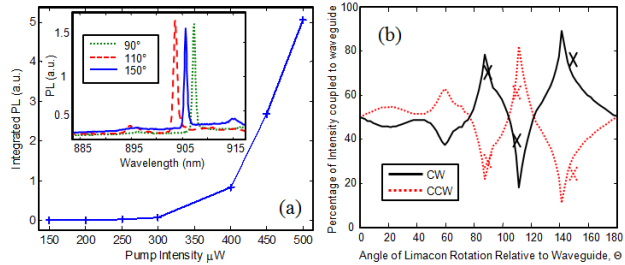

**Fig. 4.** (a) Measured emission intensity vs. pump power for the device rotated 150° relative the waveguide. Inset: Emission spectra from three identical disks with different angular rotations relative to the waveguide. (b) The percentage of the emission intensity coupled in the CW or CCW direction for the $m=12$ mode. The "×" represent to experimental data from three devices.

In conclusion, we have demonstrated directional waveguide coupling from a wavelength-scale deformed microdisk laser. Unlike a large microdisk laser that produces equal outputs in both directions of the waveguide, the wavelength-scale cavities couple the majority of the emission in a single direction of the waveguide. This behavior results from the GHS and FF effect that induce a spatial separation of the CW and CCW ray orbits. This device also reduces the spatial hole burning effect which would otherwise limit the optical gain available to the lasing mode. Although the study presented here is focused on a specific cavity shape, the results are general and applicable to a wide range of cavity shapes, provided the shape does not exhibit perfect rotational symmetry (a circular disk). We have shown that by taking advantage of wave effects that become significant in wavelength-scale cavities, it is possible to achieve functionality that is not possible in a larger device.

We thank Profs. A. Douglas Stone and Jan. Wiersig for useful discussions. This work is supported partly by NIST under the Grant No. 70NANB6H6162 and by NSF under the Grant No. DMR-0808937.